\documentstyle[times,graphics,epsfig,amsmath,amssymb,natbib,referee]{mn2e}
\def\S{\mathcal{S}}

\def\n{{\bf{\hat{n}}}}

\def\F{{\mathcal F}}
\def\u{{\bf U}}

\newcommand{\be}{\begin{equation}}
\newcommand{\e}{\end{equation}}
\newcommand{\bear}{\begin{eqnarray}}
\newcommand{\ear}{\end{eqnarray}}

\begin{document}

\title{ Cross-correlation of the  HI
  21-cm Signal and Lyman-$\alpha$ Forest: A Probe Of Cosmology}
\author[Guha Sarkar, Bharadwaj, Choudhury \& Datta ]
{Tapomoy Guha Sarkar$^{1}$\thanks{E-mail: tapomoy@cts.iitkgp.ernet.in}, 
Somnath Bharadwaj$^{1,2}$\thanks{E-mail: somnathb@iitkgp.ac.in},
Tirthankar Roy Choudhury$^{3}$\thanks{E-mail: tirth@hri.res.in} \and
Kanan K. Datta$^{4}$\thanks{E-mail: kdatta@astro.su.se}\\
$^1$Centre for Theoretical Studies, IIT, Kharagpur 721302, India\\
$^2$Department of Physics \& Meteorology, IIT, Kharagpur 721302, India\\
$^3$Harish-Chandra Research Institute, Chhatnag Road, Jhusi, Allahabad, 211019 India. \\
$^{4}$ The Oskar Klein Centre for Cosmoparticle Physics, Department of Astronomy, Stockholm University, Albanova, SE-10691 Stockholm, Sweden.
}

\maketitle
\date{\today}

\begin{abstract}
Separating the cosmological  redshifted 21-cm signal from foregrounds
is a major  challenge.  We present the cross-correlation of the
redshifted  21-cm emission  from neutral hydrogen  (HI) in the
post-reionization era with the Ly-$\alpha$ forest  as a new probe of
the large scale matter  distribution in the redshift range $z=2$ to
$3$ without the problem of foreground contamination. 
 Though the 21-cm and the Ly-$\alpha$  forest signals originate from
 different astrophysical systems, they are  
both expected to trace the underlying dark matter distribution on
large scales. The multi-frequency angular cross-correlation power
spectrum estimator 
is found to be unaffected by the discrete  quasar sampling, which only
affects the noise in the estimate.

We consider a hypothetical redshifted 21-cm observation in a single
field of view $1.3^{\circ}$ (FWHM) centered at  $z=2.2$ where the
binned 21-cm angular power spectrum can be measured  at an  SNR of
$3 \sigma$ or better across the  range $500 \le \ell \le 4000  $. Keeping
the parameters of the 21-cm observation fixed, we have estimated the
SNR for the cross-correlation signal varying the quasar angular number
density $n$ of the Ly-$\alpha$ forest survey. Assuming that the
spectra have  SNR  $\sim 5$ in pixels of length $44 \, {\rm km/s}$,  
we find that a $5 \sigma$ detection of the cross-correlation signal 
is possible  at $600\le \ell \le 2000$ with $n=4 \ {\rm deg}^{-2}$.  
This value of $n$ is well within the reach of upcoming Ly-$\alpha$
forest surveys.  
 The cross-correlation  signal will be a new, 
independent probe of the astrophysics of the diffuse IGM, the growth
of structure and the expansion history of the Universe. 
\end{abstract}

\section{Introduction.}
Observations of  the redshifted 21-cm radiation from neutral hydrogen
(HI) provides an unique opportunity for
probing the cosmological matter distribution 
over a wide range of redshifts $(0 \leq z \leq 200)$ and there
currently is considerable effort underway towards detecting this
\citep{hirev1, hirev2, hirev3}. Foregrounds from other astronomical sources which are
several orders of magnitude larger, however,  
pose a severe challenge for detecting this signal \citep{fg1, fg2,
  fg3}.
The 21-cm  emission from the post-reionization era ($z < 6$) is of
particular 
interest \citep{saini2001, poreion1, poreion2, poreion3} because the foregrounds are relatively
smaller and the HI is expected to trace the 
underlying dark matter  with a possible bias. 
These observations  hold the possibility of measuring both the matter
power spectrum and the  cosmological parameters \citep{param1, param2,
param3}. 

Interestingly, diffused HI in the intervening intergalactic medium in the same $z$  range, produces a large number of absorption lines (Lyman-$\alpha$
forest) in the spectra of distant quasars (QSO). These low neutral
density absorption lines are caused due to small baryonic fluctuations
in the IGM and has the potential to probe the matter distribution and
baryonic structure formation to very small scales.
Here the the Ly-$\alpha$ forest, whose fluctuations 
 is believed to trace the underlying dark matter, is of
special interest.
The Ly-$\alpha$ forest is known to be a valuable cosmological
probe \citep{Mandel}.  This has found a variety of applications which
include determining the matter  power spectrum \citep{pspec, pspec1, pspec2}, 
cosmological parameter estimation \citep{cosparam, cosparam1, cosparam2}, constraining the
clustering properties of dark matter on small scales \citep{viel} and
probing the reionization history  \citep{reion, reion1, reion2}. 

Though  the 21-cm emission and the Ly-$\alpha$ forest 
both originate from HI at the same $z$,   these two signals 
originate from two different kinds of astrophysical  systems. 
The Ly-$\alpha$ forest originates from small HI fluctuations
present in the primarily ionized IGM; the 21-cm emission  from these
regions is completely negligible. On the other hand, 
the bulk of the 21-cm signal originate from Damped Ly-$\alpha$
Absorbers (DLAs) which contain most of the neutral hydrogen at these
epochs \citep{xhibar, xhibar1, xhibar2}.  It is however reasonable to assume that on large
scales both these traces the same underlying dark matter, and
hence we may expect them to be correlated.

In this paper, we propose a novel probe of the large scale
matter distribution using the cross-correlation of the 21-cm
brightness  temperature  and the Ly-$\alpha$ forest transmitted flux.  The
cross-correlation signal holds the potential of independently
unveiling the same  astrophysical and  cosmological  information as
the individual auto-correlations, with the added advantage that 
the problems of foregrounds and systematics are  expected to be much
less severe for the cross-correlation. We note 
earlier studies that  consider the possibility of cross-correlating
the Ly-$\alpha$ forest  with the CMBR \citep{croft3}
and weak lensing \cite{sudeep}, and cross-correlating the
post-reionization  21-cm  signal with the CMBR \citep{tgs1} and weak
lensing \citep{tgs2}.  The cosmological 21-cm signal has recently been
detected through cross-correlations with the  6dfGRS \citep{pen09} and
the DEEP2 optical galaxy redshift survey \citep{chang10}.

\section{The  Cross-correlation Angular power Spectrum.} 
The fluctuations in the transmitted flux ${\F}(\n ,z)$  
 along a line of sight $\n$  in the 
Ly-$\alpha$ forest  may be quantified using $\delta_{\F}(\n ,z)= {\F}(\n
,z)/\bar{\F}-1$. At the large scales of interest here it
is reasonable to adopt 
the fluctuating Gunn-Peterson approximation \citep{gunnpeter, bidav,
  pspec, pspec1}
which relates the flux and the matter density contrast as 
${\F}=\exp[-A(1+\delta)^{\kappa}]$ where
$A$ and $\kappa$ are two redshift dependent functions. The function $A$
is of order unity and depends on the mean flux level, IGM
temperature, photo-ionization rate and cosmological parameters, while
$\kappa$ depends on the IGM temperature density relation \citep{mac,
  trc}. For a preliminary analytic estimate  of the  cross-correlation signal,
 we assume that $\delta_{\F}$ has been smoothed whereby  it is
 adequate to retain  only the linear term $\delta_{\F} \propto 
 \, \delta$ \citep{pspec, bidav, vielmat, slosar1} 
The higher order terms,  which have been dropped to keep the analytic
calculations tractable,  will, in principle, contribute
to the cross-correlation. We plan to address this in future studies
using simulations. 

In the redshift range of our interest ($z<3.5$) the  fluctuation in
the redshifted 21-cm brightness temperature $\delta_T(\n,z)$  traces
the underlying dark matter distribution with a possible scale
dependent bias function $b_T(k,z)$.  The bias is expected to be scale
dependent below the Jeans length-scale \citep{fang}, and  fluctuations in the
ionizing background \citep{poreion3, poreion0}  also give rise to a 
scale dependent bias.  Further, this bias is  found to grow 
monotonically with $z$ for $ 1 < z < 4 $   \citep{marin}.  However,
the simulations of \citep{bagla2}, and also \citep{poreion0} indicate
that a constant, scale independent bias is adequate at the large
scales of our interest  ( $\ell < 6000$ at  $ z \sim 2.2$).  We have
used the constant value $b_T=2$ in our analysis.

 With  these assumptions and incorporating redshift space distortions
 we may express both $\delta_{\F}$ and $\delta_{T}$   
as 
\be
\delta_{\alpha}(\n, z) =  C_{\alpha}  \int \ \frac{d^3 {\bf{k}}}{(2\pi)^3} \
e^{i {\bf{k}}.\n r}   [1 + \beta_{\alpha} \mu^2] \Delta({\bf{k}}) \,.
\label{eq:deltau}
\e
where $\alpha={\F},T$ refers to the Ly-$\alpha$ flux and  21-cm
brightness temperature respectively,
$r$ is the comoving distance, $\Delta({\bf{k}})$ is the dark matter 
density contrast in Fourier space and $\mu= {\bf \hat{ k} \cdot
  \hat{n}}$. We adopt $C_{\F}=-0.13$ and
$\beta_{\F}= 1.58$ from numerical simulations of  the Ly-$\alpha$
forest \citep{mcd03}. 

For the 21-cm we use $C_T=\bar{T} \, \bar{x}_{\rm
  HI} \,  b$  and $\beta_T=f/b$ \citep{bharad04,bali},
where 
\be
\bar{T}(z) = 4.0 \rm mK \, (1 + z )^2 \ \left(\frac{\Omega_{b0}
  h^2}{0.02} \right) \,
\left(\frac{0.7}{h} \right) \, \left(\frac{H_0}{H(z)}\right), 
\e
 $\bar{x}_{\rm HI}$ is the mean neutral hydrogen fraction,  $f$ is the linear growth parameter
of density fluctuations and $b_T$ is the bias. 
At redshifts $0 \leq z \leq 3.5$ we have $\Omega_{gas}\sim 10^{-3} $
\citep{xhibar, xhibar1} which implies that $\bar{x}_{\rm HI}= 50 \,
  \Omega_{gas}  
h^2 \, (0.02/\Omega_b h^2) = 2.45 \times 10^{-2}$ used here. As
  mentioned earlier, numerical simulations \citep{bagla} suggest that
  $b \approx 2$ which we adopt here. 

Consider next a field of view that is sufficiently small such  that
it may be treated as being flat. We may then express the unit vector
along the line of sight as  $\n = \hat{\bf{m}} + \vec{\theta} 
$,  where $\hat{\bf{m}}$ is the line of sight to the centre of the field of
view and  $ \vec{\theta}$ is a two-dimensional ($2D$)  vector on the
sky ($\theta << 1$).  In this flat sky approximation it 
 is convenient to decompose $\delta_{\F}(\vec{\theta},z)$ and
$\delta_T(\vec{\theta},z)$   into Fourier modes where we use   $\u$
as the variable conjugate to $\vec{\theta}$.  Following \citet{datta1}, 
we define the  multi-frequency angular power spectrum (MAPS) as 
\be
 P_a(\u, \Delta z) =
\frac{1}{\pi r^2} \int_0^\infty
dk_{\parallel} \ \cos(k_\parallel 
\Delta r) \ F_a(\mu) \  P(k) \,.
\label{eq:pu}
\e
Here $a=T$ refers to the HI 21-cm brightness temperature fluctuation
power spectrum  of
$\delta_T(\vec{\theta},z)$  and $\delta_T(\vec{\theta},z+\Delta z)$ at
two slightly different redshifts  $ z$ and $ z +  \Delta z$.
Similarly, $a=\F$ and $a=c$ respectively refer to the Ly-$\alpha$
forest   and $\delta_T-\delta_{\F}$ cross-correlation power spectra.
In eq. (\ref{eq:pu}),   $\Delta r$ is the radial comoving separation
corresponding to $  
\Delta z$, $P(k)$ is the dark matter power spectrum, 
$ k = \sqrt{ k_\parallel^2 + (\frac{2\pi U}{r})^2}$ and
$\mu=k_{\parallel}/k$.  The function $F_a(\mu)$ takes values 
$A^2(\mu)$ , $B^2(\mu)$ and  $A(\mu) \ B(\mu)$ corresponding to 
$a=T$, ${\F}$ and $c$ respectively. Here   $A(\mu) = C_{T}    [1 +
  \beta_T   \mu^2]$ and $B(\mu)=  C_{\F} [1 + \beta_{\F} \mu^2 ]$.  
We note that the MAPS $ P_a(\u, \Delta z)$, which is directly
related to  observable quantities, contains the entire
information of the three dimensional (3D) power spectrum through its
$U$ and $\Delta z$ dependence.

Given a field of view, it will be possible to probe $\delta_{\F}$ only along
a few, discrete lines of sight corresponding to the angular positions
of the bright quasars. We incorporate this through a sampling function 
which is a sum of Dirac delta functions 
$
\rho(\vec{\theta}) = N^{-1} \  \sum_n
\delta_D^{2}( \vec{\theta} - \vec{\theta}_n) 
$
\ where $\vec{\theta}_n$ refers to the angular positions of the quasars
and the summation extends up to $N$, the  number of quasars in the
field of view. Taking into account the discrete sampling, the observed
Ly-$\alpha$ forest flux fluctuation may be written as 
$ \delta_{{\F} o}(\vec{\theta}) = \rho(\vec{\theta}) \ \delta_{\F}(\vec{\theta}) $.
The aim here being to detect the cross-correlation power spectrum, we
define the estimator
\begin{eqnarray} 
\hat{E} (\u,\Delta z) =
\frac{1}{2 } \left[ \tilde{\delta}_{{\F} o}(\u,z) \
 \tilde{\delta}_{T}^{*}(\u,z+\Delta z)\right] \nonumber \\ + 
\frac{1}{2 } \left[ \tilde{\delta}_{{\F} o}^{*}(\u,z) \ 
  \tilde{\delta}_{T}(\u,z+\Delta z) \right] \label{eq:estimator}.
\end{eqnarray} 
where tilde  denotes the 2D Fourier transform. While it has been
assumed that the
Ly-$\alpha$ forest  and the HI 21-cm brightness
temperature  both traces the same underlying dark matter distribution,
with possibly different bias parameters, the quasars  are assumed
to be at a higher redshift and  hence  
uncorrelated with either $\F$ or $T$. 
 Using this, we have the expectation value and variance
of the estimator to be 
\be
\langle \hat{E} (\u,\Delta z)\rangle  = P_c(\u,\Delta z)
\label{eq:mean}
\e
and 
\begin{eqnarray}
\langle (\Delta \hat{E})^2 \rangle 
 =
  \frac{1}{2} P_c^2(\u,\Delta z) + \frac{1}{2} \left[ P_T(\u,0) +
N_T \right]  \nonumber  \\
 \times \left[ \frac{1}{n} \int d^2 \u' P_{\F}(\u',0) + P_{\F}(\u,0) +
   N_{\F} \right] 
\label{eq:var}
\end{eqnarray}
where $N_T$ and $N_\F$ are  the respective  noise power spectra for
$T$ and $\F$, it being assumed that the two noises are uncorrelated. 
The quasars, with  angular number density $n$, have been   assumed to be 
randomly distributed and their clustering has been ignored. We assume
that the variance $\sigma^2_{{\F} N}$ of  the pixel noise contribution
to $\delta_{\F}$ is the same across all the quasar spectra  whereby we
have $N_\F = \sigma^2_{{\F} N}/n$ for its noise power spectrum. 
The integral in eq. (\ref{eq:var}) can be simplified using
eq. (\ref{eq:pu}) to calculate $P_{\F}(\u',0)$ whereby  
\be
\int d^2\u' \ P_{\F}(\u',0) \ = \  \int \frac{d^3 {\bf
    k}}{(2 \pi)^3}  \
C_{\F}^2 [1+ \beta_{\F} \mu^2]^2 P(k)= \sigma_{\F L}^2
\e
where ${\bf k} \equiv (2 \pi \u/r,k_{\parallel})$ and $\sigma^2_{\F L}$ is the
variance of the fluctuations in the smoothed $\delta_{\F}$ arising
from the  large scale matter fluctuations and peculiar velocities. 
We have the total variance of $\delta_{\F}$ as
$\sigma^2_{\F}=\sigma^2_{\F N} + \sigma^2_{\F L}$  whereby the variance
of the cross-correlation estimator is 
\begin{eqnarray}
\langle (\Delta \hat{E})^2 \rangle   =
  \frac{1}{2}\left\{ P_c^2(\u,\Delta z) +  \left[ P_T(\u,0) +
N_T \right] \right. \nonumber \\ \left .\left[ P_{\F}(\u,0) +   \frac{\sigma^2_{\F}}{n} \right] 
\right\}
\label{eq:var1}
\end{eqnarray}
The cross-correlation signal being statistically isotropic on the sky,
we may combine estimates of the power spectrum over different
directions of $\u$ to reduce the uncertainty (or variance) in the
estimated cross-correlation signal.  Binning in $U$ and combining
estimates at different redshift values within the observational
bandwidth lead to a further reduction of the uncertainty. Finally,
incorporating the possibility of observations in several independent
fields of view, we use $N_E$ to denote the total number of independent
estimates that are combined. The uncertainty or noise in the resulting
combined estimate of $P_c(U,\Delta z)$ is 
$\sigma^2 \  = \  \langle (\Delta  \hat{E})^2 \rangle / {N_E} \,$.
It is convenient to  express our results in terms of the angular
 multipole $\ell= \  2 \pi U$. 
 We then have 
$ N_E=(\ell + \frac{1}{2})\,  \Delta \ell \, (B/\Delta \nu) \, f \, N_F$ where
 $\Delta \ell$ is the width of the $\ell$ bin, $B$ the frequency
 bandwidth of the 21-cm observation, $\Delta \nu$ the frequency interval
 beyond which we have an 
 independent estimate of the signal, $f$ the fraction of the sky
 covered by a single field of view and $N_F$ the number of independent
 fields of view  that are observed. 

\begin{figure}
\begin{center}
 \mbox{\epsfig{file=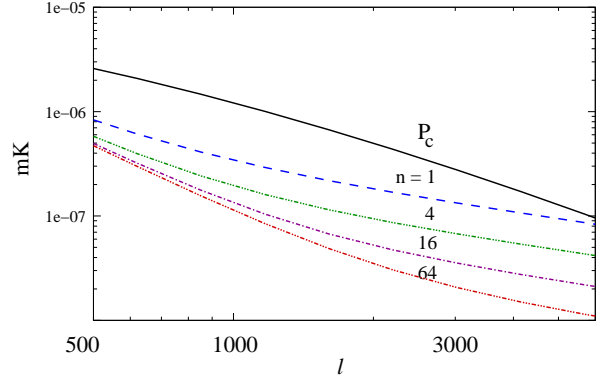,width=0.4750\textwidth,angle=0}}
\caption{The magnitude of the cross-correlation  angular power
  spectrum (topmost curve)   and  $1 \sigma$ errors (lower curves) for   $z=2.2$. Note that
  $\delta_{\F}$ and $\delta_T$ are anti-correlated.} 
\label{fig:cross}
\end{center}
\end{figure}

\begin{figure}
\begin{center}
 \mbox{\epsfig{file=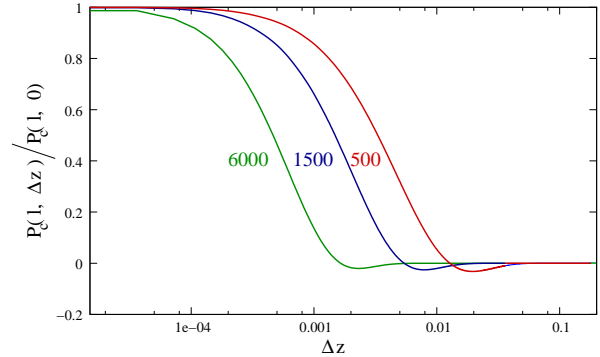,width=0.475\textwidth,angle=0}}
\caption{The decorrelation of $P_c(\ell,\Delta z)$  with increasing
 $\Delta z$ for the  representative $\ell$ values  shown in the figure.}  
\label{fig:decorrelation}
\end{center}
\end{figure}

\begin{figure}
\begin{center}
 \mbox{\epsfig{file=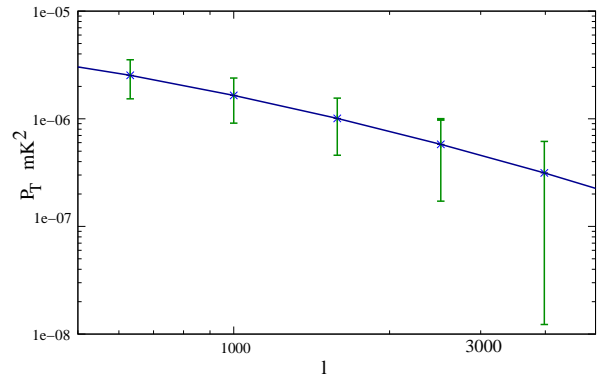,width=0.475\textwidth,angle=0}}
\caption{The HI 21-cm angular power spectrum  for  $z =   2.2$ with $3 \sigma$ error bars}
\label{fig:hi}
\end{center}
\end{figure}

\section{Detectability.}

We next  estimate  the survey parameters that will be required to 
detect the cross-correlation signal. It is, in principle, possible to
vary the parameters of both the redshifted  21-cm survey and the
Ly-$\alpha$ survey. To keep the analysis  simple we
restrict our attention  to a situation where the parameters of the
redshifted 21-cm survey are fixed, and vary the parameters of only the
Ly-$\alpha$ survey.

 The quasar  distribution  is known to peak between $z=2$ and
$3$. For any particular quasar, it is possible to reliably estimate
$\delta_{\F}$   in a small redshift range close to the quasar's redshift.   
The region very close to the quasar is excluded due to the quasar's
Stromgren sphere, and  large redshift separations are excluded to
avoid Ly-$\beta$ contamination.  Based on this we have only
considered quasars in the $z$ range $2.2-3.0$ for our estimates, and
we have chosen a region  centered at $z=2.2$ for our estimates.

The predicted HI- $\F$ cross correlation angular power spectrum is
shown (Figures  \ref{fig:cross} and \ref{fig:decorrelation}) 
  assuming  cosmological parameters from   WMAP $5$
  results \citep{komatsu}. The $\ell$ dependence closely follows that
  of the HI angular power spectrum (Figure \ref{fig:hi}). The
signals  at two different redshifts  $z$ and $z+ \Delta z$,
we find, decorrelate  rapidly with increasing $\Delta z$, the decline
being faster at larger $\ell$ values.

The currently functioning GMRT \citep{swarup} can, in
principle be used to probe the redshifted HI 21-cm signal all the way
from $z\sim0$ to $z \sim 8$ \citep{bali}. The GMRT, at present,
has neither the exact frequency band nor the desired sensitivity for
the proposed observation. In principle it would   not be very 
difficult, in future, to cover the required frequency and increase 
the number of  antennas to increase the  sensitivity. For the present
analysis we consider a hypothetical array, possibly an extended
version of the GMRT or some other  future radio telescope with
60 antennas   similar to the GMRT, distributed randomly over a 1 km
$\times$ 1 km square. Note that this is roughly $4$ times the number
of antennas currently available in the GMRT central square, and
henceforth we refer to this as the extended  GMRT (EGMRT).  Each
antenna is $45 \, {\rm m}$ in diameter, with a field of view 
$1.3^{\circ}$ (FWHM), total system temperature $100 \, {\rm K}$ and
antenna gain $0.33 \, {\rm K/Jy}$. We assume that the observations
are carried out over a frequency band of $32 \, {\rm MHz}$ centered at
$430 \, {\rm MHz}$ using channels of width $62.5 \, {\rm kHz}$ each.  
The frequency separation $\Delta \nu$ over which the 21-cm signal
remains correlated roughly scales as $\Delta \nu = 1 \, {\rm MHz} \, 
(\ell/100)^{-0.7}$ \citep{poreion7}. We assume that the signal is
averaged over frequency bins of this width  to increase the
SNR.  Considering $1,000$ hrs of observation in a single field of 
view, a $3 \ \sigma$ (or better) measurement of the HI 21-cm power spectrum
will be possible at $\ell \le 4000$ (Figure \ref{fig:hi}). Note that
the error is dominated by the system noise, the cosmic variance being
considerably smaller. This justifies why we have considered
observations in a single field of view instead of distributing the
$1000 \, {\rm hr}$ observation over different fields. 

Considering next the Ly-$\alpha$ forest surveys, we note that these
typically cover a much larger angular region and redshift interval
compared to the 21-cm observation that we have considered. For
example, the  SDSS  Data Release 3 \citep{sneider}, whose data is
currently available, covers  $4,188 \ {\rm  deg}^2$ of the sky.
The number density of  quasars in the redshift range $z=2.2 - 3.0$ is 
$n= 1.0 \ {\rm   deg}^{-2}$ for this survey.  The cross-correlation is
restricted to the angular extent of the 21-cm observation and the
redshift interval $\Delta z= 0.24$ centered at $z=2.2$ which
corresponds to a bandwidth of $32 \, {\rm MHz}$ centered at $430 \,
{\rm MHz}$. The channel width $62.5 \, {\rm kHz}$ of the 21-cm
observations  corresponds to $\Delta z=4.7 \times 10^{-4}$ or
equivalently $v_{\parallel}=44 \, {\rm km/s}$.

The analysis of fluctuations in the Lyman-$\alpha$ forest
\citep{dodorico, coppolani} show that the variance has a value 
$\sigma^2_{\F L} \approx 0.02$ for  $\delta_{\F}$ smoothed over
$\sim 50 \, {\rm km/s}$ along the line of sight. This smoothing
 is  comparable to the channel width of the 21-cm observations, 
and for simplicity we assume that Ly-$\alpha$ pixel length is 
exactly the same as the 21-cm channel width. Note that this 
is  smaller than the typical  $\Delta z$ value where the signal 
 decorrelates (Figure  \ref{fig:decorrelation}).    For the pixel
 noise contribution, we assume that $S/N=5$ for every pixel of the
 spectra used to estimate the cross-correlation. This gives
 $\sigma^2_{\F N}=0.04$, whereby  $\sigma^2_{\F}=0.06$ for pixels 
of length $44 \, {\rm km/s}$ or $62.5 \, {\rm kHz}$. 
As noted earlier, it is advantageous to average the signal over an
interval $\Delta \nu = 1 \, {\rm MHz} \, (\ell/100)^{-0.7} \ge 62.5 \, 
{\rm kHz}$  before correlating. The value of $\sigma^2_{\F}$ will come
down due to this averaging. Assuming that the pixel noise in different
pixels is uncorrelated, we have $\sigma^2_{\F N} = 0.04 \ (62.5 \, {\rm
  kHz}/\Delta \nu)$. Analysis of the line of sight correlation
function of $\delta_{\F}$ indicate that we may expect $\sigma^2_{\F L}$
to scale faster than $(\Delta \nu)^{-1}$. For the purpose of this
paper we assume that both $\sigma^2_{\F L}$ and $\sigma^2_{\F N}$ have
the same scaling whereby $\sigma^2_{\F } = 0.06 \ (62.5 \, {\rm
  kHz}/\Delta \nu)$. which we use in eq. (\ref{eq:var1})  for our
noise estimates. The error introduced by the last assumption, will at
worst, cause the noise for the cross-correlation signal to be
over-estimated. 

We present noise estimates (Figure  \ref{fig:cross}) considering  
quasar  angular number densities $ n = 1, 4, \ 16$  and $ 64 \
{\rm   deg}^{-2}$.  While our intention is primarily to estimate the
quasar number density that will be required to detect the
cross-correlation signal, we note that the $n$ values chosen are viable
with existing or future surveys. The currently available 
SDSS \citep{sneider} has $ n \sim  1 {\rm deg}^{-2}$  and the upcoming 
  BOSS\footnote{http://cosmology.lbl.gov/BOSS/}\citep{mcd1}  is
  expected to have $ n \sim  16 {\rm deg}^{-2}$, while 
the proposed future   BIGBOSS \citep{bigboss} is anticipated to reach
$ n >  64 {\rm     deg}^{-2}$. We find that a $3\sigma$ and $5\sigma$
detection will   be possible at  $\ell \le 2000$ for $n=1$ and $4 \
{\rm deg}^{-2}$   respectively. A $5\sigma$ (or better) detection will
be possible over the entire $\ell$ range for $n=16$ and $64 \, {\rm
  deg}^{-2}$. There is a further reduction of  noise by a factor $
N_F^{-1/2}$ if the same observation is repeated in  multiple fields of
view.

Unlike the $\delta_{\F}$ auto-correlation power spectrum which is
Poisson noise dominated,  the cross-correlation signal itself is not
affected  by the discrete quasar sampling. However its variance is
very sensitive to this, and a dense quasar sampling  will allow the
cross-correlation to be measured at a high level of precision.

 The discussion so far has completely bypassed several
  observational difficulties which pose a severe
  challenge. Considering first the Ly-$\alpha$ forest, errors in
  continuum  fitting and  subtraction would result in  an additive
error in the estimated  $\delta_{\F}$ which will inhibit recovery of
the underlying power spectrum at large scales.  \citet{pspec4} and
  \citet{mcd06}   have studied  this   issue extensively and have
  proposed several   techniques to mitigate   the  contribution from
  such errors. While  an additive error in $\delta_{\F}$ could have
  severe repercussions   for the   large-scale  power spectrum    
 estimated from  the   auto-correlation of   $\delta_{\F}$
  \citep{kim04},  we do not 
 expect these errors to be correlated with the  21-cm data. An
  additive error in 
  $\delta_{\F}$ will manifest itself as an extra contribution to the
  noise for the cross-correlation power spectrum (eq. \ref{eq:var1})
  which, in turn may degrade the SNR and hence affect the
  detectability of the cross-correlation signal.

The  redshifted 21-cm  signal is    buried under foregrounds which are
  several orders of magnitude   larger \citep{shaver99,fg4,fg1,fg6,
  fg3,bernardi09,pen09,fg10}. Extragalactic point sources and the
  diffuse synchrotron radiation from our own Galaxy are the  two most
  dominant foreground components.   The free-free emissions from our
  Galaxy and external galaxies  make much smaller
  contributions  \citep{shaver99}, though each of  these  is
  individually   larger than the HI signal.   Several different
  techniques 
  have been proposed for  separating the 21-cm signal from the
  foregrounds. All of these depend on the fact that the foregrounds
  are expected to have a   continuum frequency spectrum, and their
  contribution at two different frequencies separated by $\Delta \nu$
  is expected to be correlated well beyond $\Delta \nu \sim 5 \, {\rm
  MHz}$ . The 21-cm signal, however, is predicted to decorrelate
  within $\Delta \nu \sim 1 \, {\rm MHz}$ for angular scales of our
  interest \citep{poreion1}. A possible technique for foreground
  removal is to subtract out 
 any smooth  frequency dependent
  component  either from the image cube  \citep{jelic, fg7, fg8}  or 
from the gridded  visibilities \citep{fg9}. Another possible approach
  is to first estimate the  multi-frequency angular power  spectrum
of the radio-interferometric data   and then subtract out any
  component that remains correlated over large frequency separations 
\citep{fg3,fg10}. 

The foregrounds of the redshifted 21-cm signal are expected to be
uncorrelated with the Ly-$\alpha$ forest $\delta_{\F}$ and also any
errors arising in it from continuum subtraction. We  do not
expect the foregrounds to contribute to the estimated
cross-correlation signal, and we anticipate that  the problem of
foreground removal will be  considerably less severe as compared to
the auto-correlation. Errors in foreground subtraction will manifest
as an extra source of noise for the  cross-correlation signal. The
fact that the 21-cm $\delta_T$ and the Ly-$\alpha$ $\delta_{\F}$ at
two different redshifts separated by $\Delta z$ decorrelate rapidly as
$\Delta z$ is increased (Figure \ref{fig:decorrelation}) should help in
identifying any foreground contamination. 

Errors in calibrating the radio observations is another possible
source of uncertainty in the 21-cm signal. This will lead to errors in
the overall amplitude of the cross-correlation signal, or equivalently
contribute to uncertainties in estimates of the quantity $C_{\F} C_T$
defined  in Section 2.

In conclusion, we propose the 21-cm and Ly-$\alpha$ forest
cross-correlation signal as a  tool to  measure the  large-scale
matter distribution. The problem of foreground removal is expected to
be considerable less severe for the cross-correlation than for the 
21-cm auto correlation signal.The cross-correlation signal  will probe a
variety of issues like   the astrophysics of the diffuse IGM, the
growth of large-scale  structures and the expansion history of the
Universe.

\bibliography{references}

\end{document}